Loss-induced enhanced transmission in anisotropic density-near-zero acoustic metamaterials

Chen Shen and Yun Jing[a]

Department of Mechanical and Aerospace Engineering, North Carolina State University,

Raleigh, North Carolina 27695, USA

[a]yjing2@ncsu.edu

Abstract:

Anisotropic density-near-zero (ADNZ) acoustic metamaterials are investigated theoretically and numerically in this letter and are shown to exhibit extraordinary transmission enhancement when material loss is induced. The enhanced transmission is due to the enhanced propagating and evanescent wave modes inside the ADNZ medium thanks to the interplay of near-zero density, material loss, and high wave impedance matching in the propagation direction. The equi-frequency contour (EFC) is used to reveal whether the propagating wave mode is allowed in ADNZ metamaterials. Numerical simulations based on plate-type acoustic metamaterials with different material losses were performed to demonstrate collimation enabled by the induced loss in ADNZ media. This work provides a different way for manipulating acoustic waves.

In recent years, acoustic metamaterials with exotic constitutive parameters have received considerable interest due to their unique features in manipulating acoustic waves[1–6]. Numerous novel applications have been realized by acoustic metamaterials, including acoustic cloaking[7,8], hyperlensing[9], and energy funneling[10]. Recently, following the concept of epsilon-near-zero (ENZ) metamaterials in electromagnetic waves, density-near-zero (DNZ) metamaterials have been proposed in the acoustic regime, showing exciting potential for controlling acoustic waves[11–14]. For example, subwavelength imaging and extraordinary transmission using DNZ metamaterials have been proposed[15,16]. In ENZ metamaterials, material loss has been demonstrated to introduce transparency, omni-directional collimation, and counterintuitively, improved transmission[17,18]. However, little work has been done to investigate the effect of material loss in acoustic ADNZ metamaterials, even though great potential of manipulating acoustic waves using ADNZ metamaterials has been suggested[13,15]. In real world applications, material loss is inevitable and its effect on ADNZ metamaterials should be examined. In addition, as we will show in this paper, loss induced in acoustic ADNZ metamaterials may lead to applications such as collimation and subwavelength imaging. Inspired by the study on ENZ metamaterials, this letter will examine the effect of material loss in ADNZ metamaterials where only one component of the mass density tensor is close to zero. The enhanced transmission and collimation effect of ADNZ acoustic metamaterial induced by material loss will be demonstrated. This letter shows that when acoustic waves reach an ADNZ metamaterial slab from a certain incident angle, they will bend and be collimated towards the normal direction when a certain amount of material loss is present. The underlying mechanism is discussed, which reveals that high impedance matching condition can be realized by material loss. These effects are verified by the EFC analysis and full wave numerical simulations based on real structures exhibiting anisotropic density-near-zero property.

Consider a homogeneous ADNZ medium whose effective density is positive in the x-direction and near-zero in the y-direction, i.e., $\rho_x > 0$ and $\rho_y \to 0$ [19]. The general dispersion relation for a two-dimensional scenario reads

$$\frac{k_x^2}{\rho_x} + \frac{k_y^2}{\rho_y} = \frac{\omega^2}{B}, \tag{1}$$

where $k$ and $\omega$ are the wave number and angular frequency, respectively, $B$ is the scalar bulk modulus. To include material loss in the ADNZ medium, $\rho_y$ is considered a complex number having the form $\rho_y = \text{Re}(\rho_y) + j\,\text{Im}(\rho_y)$. Without losing generality, the background medium is assumed to be air with density $\rho_0$, and $\rho_x = \rho_0 = 1.2\,\text{kg/m}^3$. The EFCs of three different cases are plotted in Fig. 1 based on Eq. (1), with the same $\text{Re}(\rho_y) = 0.02$ and different losses $\text{Im}(\rho_y) = -0.2$, $\text{Im}(\rho_y) = -2$, and $\text{Im}(\rho_y) = -20$. The sign of the imaginary part of the density depends on whether $e^{j\omega t}$ or $e^{-j\omega t}$ is used. The principle is that the sign should be chosen so that wave decays. These three values correspond to low loss, moderate loss, and high loss, respectively. The EFC of the ADNZ medium is represented by a solid curve and the EFC of free space is represented by a dashed curve for comparison. It can be seen from Fig. 1 that the general EFC of the lossy media is an ellipsoid. When the material loss increases, the major axis of the ellipsoid changes from the horizontal axis to the vertical axis. For an incoming plane wave with incident wave vector $k_i$, incident and transmitted waves should have the identical y-component wave vector due to the conservation of momentum[20,21]. For a 30° incident beam, it is indicated by Fig. 1 that, for the low loss case, the corresponding $k_x$ inside ADNZ medium has no solutions since no $k_y$ component can be found on the EFC matching the free space $k_{0y}$ component. However, for the moderate and high loss cases, $k_x$ exists, indicating wave propagation is allowed in the medium. As the group velocity $v_g$ must lie normal to the EFC, for the high loss case, the transmitted wave vector is pointing in the x-direction since the EFC is almost flat. Therefore, the transmitted energy is collimated towards the normal direction if a large material loss is introduced.

We further investigate the transmission characteristics of the ADNZ slab with various losses. For both propagating and evanescent waves transmitting through an anisotropic layer, the transmission coefficient is given as[15]:

$$T = \frac{4Z_x Z_{0x} e^{-jk_x L}}{(Z_x + Z_{0x})^2 - (Z_x - Z_{0x})^2 e^{-2jk_x L}}, \tag{2}$$

where $Z_x = \omega \rho_x / k_x$ and $Z_{0x} = \omega \rho_0 / k_{0x}$ are the wave impedances, $L$ is the thickness of the ADNZ slab (that is, along the x-direction), $k_{0x} = \sqrt{k_0^2 - k_{0y}^2}$ and $k_x$ are x-component wave vectors in the free space and anisotropic medium, respectively. Note that $k_x$ can be calculated from $\rho_y$ and $\rho_x$ via Eq. 3. In the y-direction, the slab is assumed to be infinitely long and $\text{Re}(\rho_y)$ is assumed to be 0.02. The imaginary part of $\rho_y$, which reflects different losses from a homogenized medium perspective, will be included in the effective density to calculate the transmission coefficients from Eq. (2). Figure 2(a) shows the transmission coefficients of both propagating and evanescent waves at 2545 Hz for $L = 30$ cm with effective densities abovementioned used. It can be seen that for normal incidence ($k_{0y} = 0$), the transmission coefficients are high in all three cases. This is because the impedance in the x-direction $Z_x$ matches with the free space impedance $Z_0$, as $k_{0y}$ is zero. For oblique incidences, especially for large incident angles (large $k_{0y}$ values), the transmission coefficients for both propagating and evanescent waves increase significantly when more loss is induced. For the low loss case, the transmission coefficient quickly drops to very small values as $k_{0y}$ increases, implying no propagating mode is allowed in the medium. Note that as illustrated in Fig. 1, sound energy mainly propagates in the x-direction for a large material loss. We hereby evaluate the value of $Z_x$ in terms of $\text{Im}(\rho_y)$. Since at the interface of the ADNZ medium and free space, we have $k_{0y} = k_y$ due to conservation of momentum[21], the wave vector $k_x$ can be obtained by:

$$k_x = \sqrt{\rho_x \left( \frac{\omega^2}{B} - \frac{k_y^2}{\rho_y} \right)}. \tag{3}$$

The wave impedance $Z_x$ is thus:

$$Z_x = \omega \sqrt{\rho_x / \left( \frac{\omega^2}{B} - \frac{k_y^2}{\rho_y} \right)}. \tag{4}$$

By inserting complex $\rho_y$ into Eq. (4), the dependence of $Z_x$ on $\text{Im}(\rho_y)$ is shown in Fig. 2(b). As a reference, the free space impedance $Z_0 = \rho_0 c_0$ is also included in the figure. It can be clearly observed that for large material losses, the impedance matching condition is fulfilled within a broad range of $k_{0y}$ starting from 0. These results are not surprising, as we can observe higher transmission in Fig. 2(a) for a certain $k_{0y}$ value with large losses. We also note that the enhanced transmission when $k_{0y} > 1$ implies that the evanescent components can be coupled through an ADNZ slab and is favorable for subwavelength imaging applications[22].

To verify the proposed phenomenon with numerical simulations, we construct the ADNZ medium utilizing plate-type acoustic metamaterials, which yield negative density below a cut-off frequency and a near-zero-density around the cut-off frequency (i.e., the first resonance frequency of the plate)[19]. The two-dimensional (2D) ADNZ metamaterial is depicted in Fig. 3(a), where periodically arranged, clamped square plates facing the y-direction are placed inside a 2D waveguide. The plate is assumed to be made of paper and has the same material property as that in an earlier study[19]. Different material losses are now considered to give rise to complex effective density in the y-direction. The Poisson's ratio, density, width and thickness for the plate are 0.33, 591 kg/m$^3$, 20 mm and 0.3 mm, respectively. Since wave propagation inside this structure can be decoupled in the x- and y-directions[19], the effective properties in these two directions can be estimated separately. To this end, one-dimensional (1D) models are first utilized to study the effective density of the ADNZ metamaterial. Finite element package COMSOL MULTIPHYSICS is adopted for numerical simulations. Since there are no plates in the x-direction, the effective density along the x-direction is considered as that of air[19], and no loss is considered, i.e., $\rho_x = 1.2$ kg/m$^3$. The setup for evaluating the complex effective density of the ADNZ metamaterial in the y-direction is shown in Fig.3 (b), where periodically arranged plates are placed in a square waveguide with width $a = 20$ mm. Each plate is separated by a distance $d = 20$ mm. To include material loss in the simulations, the Young's moduli of the plates are set to be complex numbers, i.e., $E = \text{Re}[E] + j\,\text{Im}[E]$. The same real parts of the Young's moduli are used, i.e., $\text{Re}[E] = 2.61\,\text{GPa}$. Three values are chosen for the imaginary parts: $\text{Im}[E] = 0.0261\,\text{GPa}$, $\text{Im}[E] = 0.261\,\text{GPa}$ and $\text{Im}[E] = 2.61\,\text{GPa}$. The corresponding loss factors ($\tan\delta$) are 0.01, 0.1, and 1. They correspond to low, moderate and high loss case, respectively. As can be seen below, the complex Young's modulus will translate to the complex effective density via the homogenization process.

A lumped model is used to estimate the effective density in the y-direction, with expression $\rho_y = \frac{Z_{AM}}{j\omega} \cdot \frac{1}{dA}$ [19], where $Z_{AM}$ is the acoustic impedance of the plate, $A = a^2$ is the cross-sectional area of the waveguide. The acoustic impedance of the square plate used in the simulation is calculated as $Z_{AM} = \frac{Z_m}{A^2} = \frac{\iint \Delta p A}{j\omega \xi A^2}$. The pressure difference across the plate $\Delta p$, and the average transverse displacement of the plate $\xi$ are estimated numerically using COMSOL in order to get the mechanical impedance of the plate $Z_m$. Figure 4 depicts the calculated effective density of the three cases. Since the real part is loss-independent, it is only represented by a single curve. It can be found from Fig. 4 that with higher material loss, the absolute value of the imaginary part of $\rho_y$ increases. The real part of $\rho_y$ is near zero ($\text{Re}(\rho_y) = 0.02$) around 2545 Hz and the corresponding $\text{Im}(\rho_y)$ is -0.2, -1.9, and -18.8, respectively.

Full wave simulations based on effective medium and real structures are carried out to study the enhanced transmission for loss-induced ADNZ metamaterials. A Gaussian beam with frequency 2545 Hz is transmitted with an incident angle 30° to the ADNZ metamaterial slab. The corresponding acoustic pressure and intensity fields are plotted in Fig. 5. It is clear that for the low loss case, the incident wave cannot excite propagating modes inside the ADNZ medium and the acoustic energy vanishes quickly inside the slab. When the loss increases, higher transmission is observed, and the acoustic energy is collimated in the x-direction, which is well predicted by the theory presented above. This seems to be counterintuitive, as more material losses increase the transmission. However, a rigorous analysis of the EFC shows that in the high loss case, the acoustic waves are forced to travel along the x-direction (Fig. 1(c)), where there are no plates and therefore no energy loss in that direction (Fig. 5(c)). This is consistent with the fact that the homogenized acoustic medium does not yield loss in the x-direction.

To conclude, we have studied the enhanced transmission phenomenon in loss-induced ADNZ acoustic metamaterials. Theoretical analysis revealed that the enhanced transmission is due to the enhanced propagating and evanescent wave modes and better matched wave impedance. The collimation effect can be realized by introducing a large material loss. Numerical simulations were conducted to verify the theory using plate-type acoustic metamaterials. The findings in this paper may find applications in directional sensing and acoustic imaging.

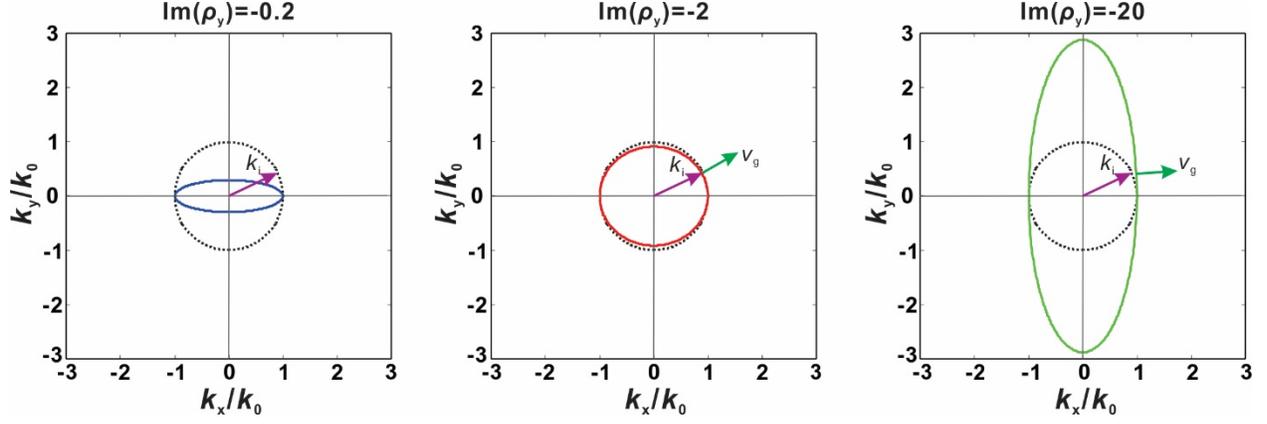

FIG. 1. EFC of an ADNZ medium with low, moderate and high losses. Dashed line represents the free space. The incident wave vector $k_i$ has a 30° incident angle.

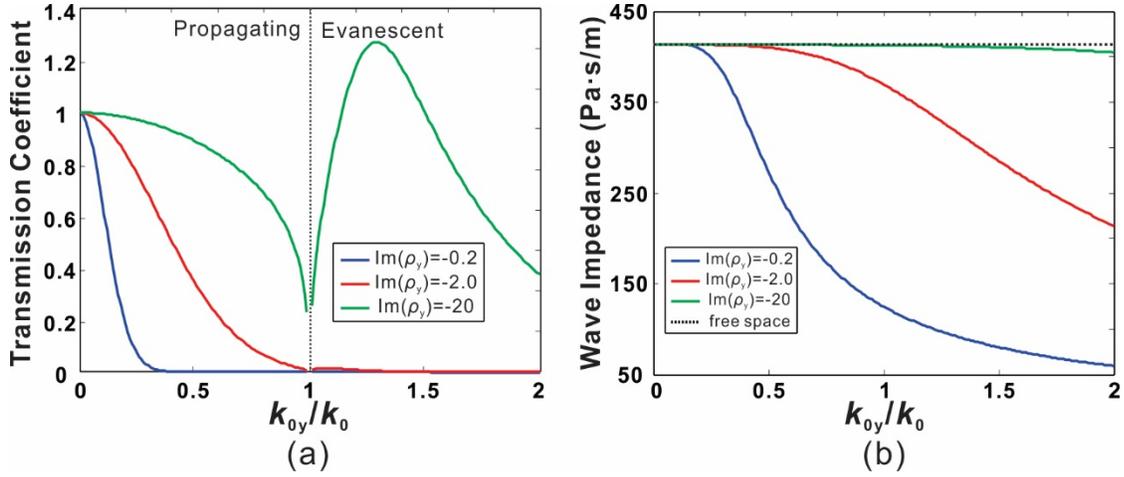

FIG. 2. (a) Transmission coefficients of the ADNZ medium with various losses. (b) The x-direction wave impedance of the ADNZ medium with various losses.

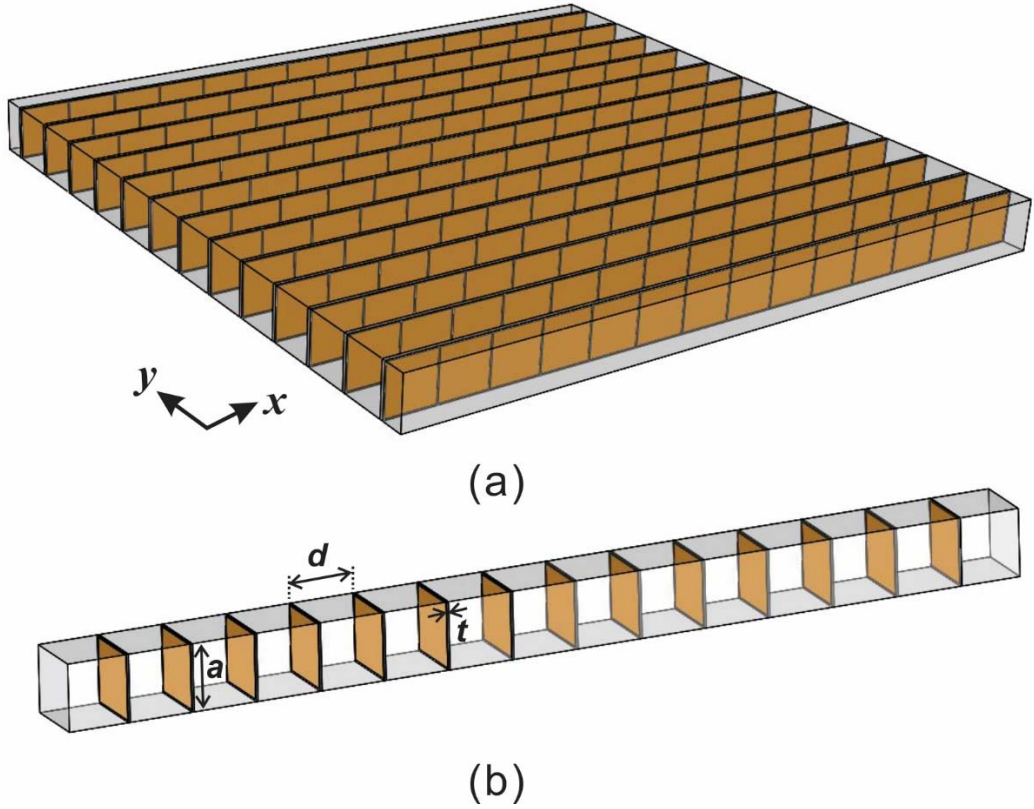

FIG. 3. (a) 2D ADNZ metamaterial (b) y-direction of the ADNZ medium, where the real part of the effective density is close to zero around the first resonance frequency of the plate. The imaginary part is tuned by the material loss.

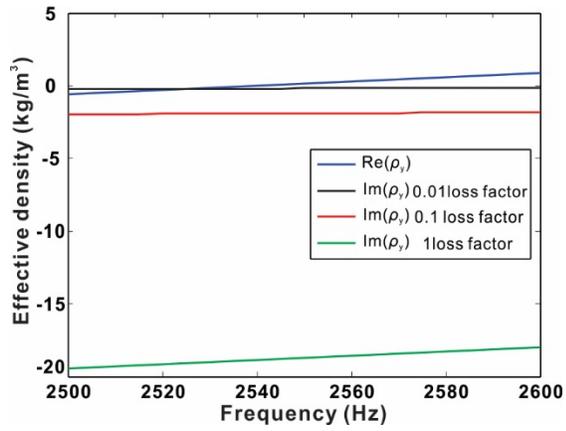

FIG. 4. Effective density along the y-direction with different losses.

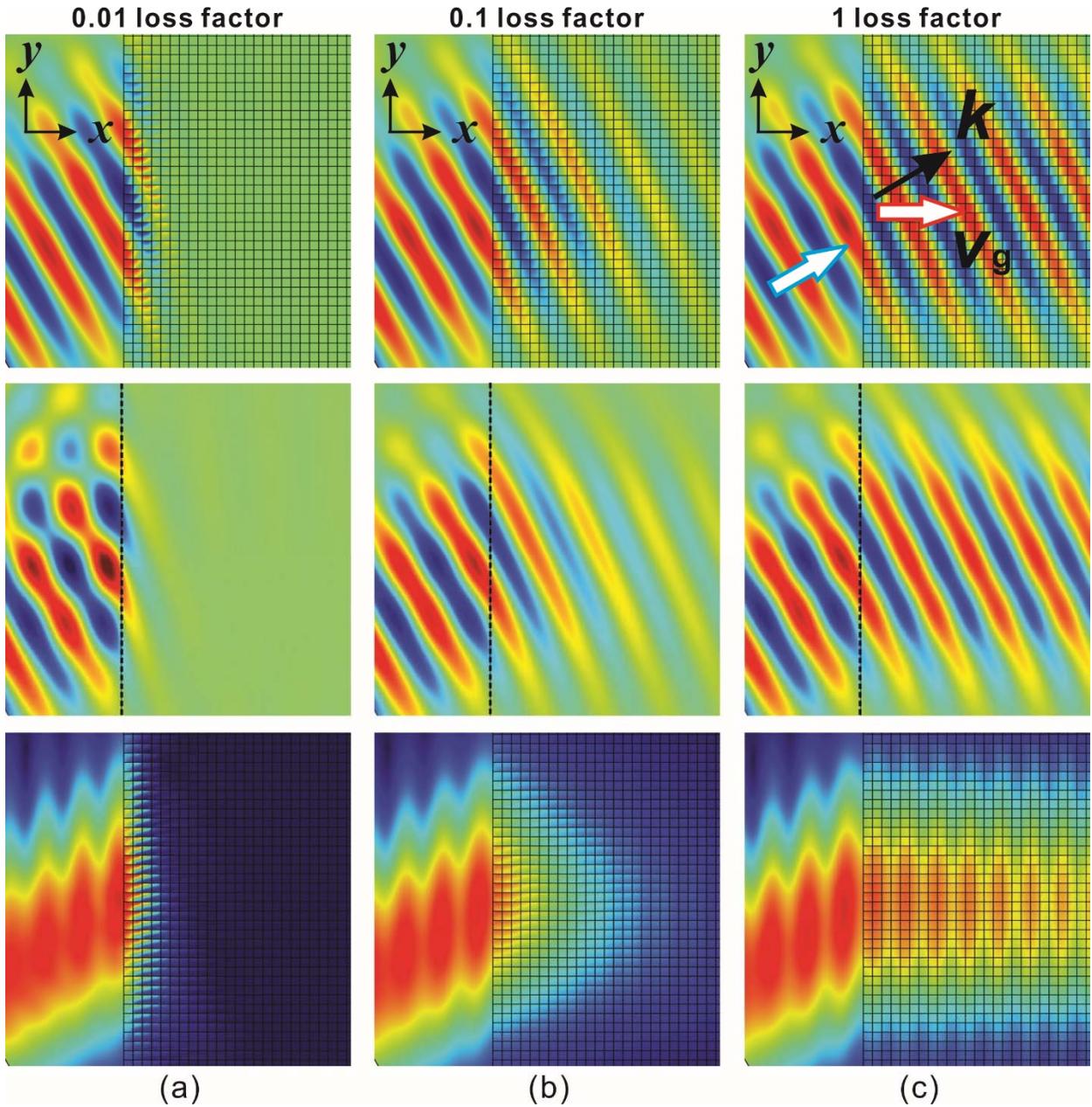

FIG. 5. The acoustic pressure and intensity fields. Top: acoustic pressure fields using real structure. Middle: acoustic pressure fields using effective medium, dotted lines denote the interface of free space and ADNZ medium. Bottom: acoustic intensity fields. (a) Low loss. (b) Moderate loss. (c) High loss. Collimation and enhanced transmission are clearly observable in the high loss case.